\documentstyle{elsart}

\newcommand{\hw}{D$_{2}$O}
\newcommand{\ptg}{$^{3}$H($p,\gamma$)$^{4}$He}

\newcommand{\pt}{$pT$}
\newcommand{\dtn}{$^{2}$H(t,n)$^{4}$He}
\newcommand{\tdn}{$^{3}$H(d,n)$^{4}$He}
\newcommand{\ttnn}{$^{3}$H(t,nn)$^{4}$He}

\newcommand{\detone}{D$_{135}$}
\newcommand{\dettwo}{D$_{90}$}
\newcommand{\detthree}{D$_{45}$}
\newcommand{\baftwo}{BaF$_{2}$}

\input epsf

\begin{document}
\begin{frontmatter}

\title{A Compact $^{3}$H(p,$\gamma$)$^{4}$He 19.8-MeV Gamma-Ray Source
for Energy Calibration at the Sudbury Neutrino Observatory}
  
\author{A.W.P. Poon\thanksref{UW}\thanksref{LBL},}
\author{R.J. Komar, C.E. Waltham}
\address{Department of Physics and Astronomy, The University of 
British Columbia, Vancouver, BC, Canada V6T 1Z1}

\author{M.C. Browne\thanksref{NCSU}\thanksref{LANL},}
\author{R.G.H. Robertson}
\address{Nuclear Physics Laboratory, University of Washington, Seattle, WA 98195, USA}

\author{N.P. Kherani}
\address{Ontario Hydro Technologies, 800 Kipling Avenue, Toronto, ON, Canada M8Z 5S4}

\author{H.B. Mak}
\address{Department of Physics, Queen's University, Kingston, ON, Canada K7L 3N6}

\thanks[UW]{Also affiliated with Nuclear Physics Laboratory, 
University of Washington, Seattle, WA 98195, USA} 

\thanks[LBL]{Present and Corresponding address: Institute for Nuclear and Particle 
Astrophysics, Nuclear Science Division, Lawrence Berkeley National 
Laboratory, Mail Stop 50-208, Berkeley, CA 94720, USA.  Tel.: (510) 495-2467; Fax: 
(510) 486-4738; E-mail: AWPoon@lbl.gov.}

\thanks[NCSU]{Also affiliated with Department of Physics, North 
Carolina State University, Raleigh, NC 27695, USA }

\thanks[LANL]{Present address: NIS-5, Los Alamos National Laboratory, Los 
Alamos, NM 87545, USA}
    
\begin{abstract}

The Sudbury Neutrino Observatory (SNO) is a new 1000-tonne \hw\ 
\v{C}erenkov solar neutrino detector.  A high energy gamma-ray source is needed to 
calibrate SNO beyond the $^8$B solar neutrino endpoint of 15~MeV. 
This paper describes the design and construction of a source that 
generates 19.8-MeV gamma rays using the \ptg\  reaction (``\pt''), and 
demonstrates that the source meets all the physical, operational and 
lifetime requirements for calibrating SNO. An ion source was built 
into this unit to generate and to accelerate protons up to 30~keV, and 
a high purity scandium tritide target with a scandium-tritium atomic 
ratio of 1:2.0$\pm$0.2 was included.  This \pt\ source 
is the first self-contained, compact, and portable high energy 
gamma-ray source ($E_\gamma>$10~MeV).

\end{abstract}
\end{frontmatter}

\section{Introduction}
\label{sec:introduction}

The Sudbury Neutrino Observatory (SNO)~\cite{sno_paper} is a new heavy 
water (\hw) \v{C}erenkov solar neutrino detector.  The detector is 
unique in its use of 1000 tonnes of \hw\ as target, which allows the 
detection of electron neutrinos and neutrinos of all active flavours through the following 
channels:
\begin{eqnarray}
\label{eq:cc}
\nu_e + d     & \rightarrow & p + p + e^- -1.44 \mbox{ MeV}  \\
\label{eq:nc}
\nu_x + d     & \rightarrow & p + n +\nu_x-2.22 \mbox{ MeV}  \\ 
\label{eq:es}
\nu_x + e^-   & \rightarrow & \nu_x + e^- 
\end{eqnarray}
This ability to measure the total flux of all active flavours of 
neutrinos originating from the Sun will allow SNO to make a 
model-independent test of the neutrino oscillation hypothesis.  

The SNO collaboration needs a high energy calibration point beyond the
$^{8}$B solar neutrino energy endpoint of $\sim$15~MeV. This
calibration point is very important in understanding the detector's
energy response because \v{C}erenkov light production is not exactly linear in energy (e.g.
energy loss to low energy electrons below the C\v{e}renkov threshold). 
As the energy increases, the probability that a photomultiplier tube
would get hit by more than one \v{C}erenkov photon
increases.  Therefore, a calibration point beyond the solar neutrino
energy endpoint will provide vital information on this multiple hit
effect at energies beyond the solar neutrino endpoint.

In the arsenal of calibration sources at SNO, the ``\pt'' source, 
which employs the \ptg\ reaction to generate 19.8-MeV gamma rays, has 
the highest energy.  This \pt\ source is the first self-contained, 
compact, and portable high energy gamma-ray source ($E_\gamma>$10~MeV).

In this paper various aspects of the construction and operation of the
\pt\ source are described.  In Section~\ref{sec:design_criteria} the
design criteria for a high energy gamma-ray calibration source at SNO
are outlined.  Attributes of the \ptg\ reaction are discussed in
Section~\ref{sec:attributes}.  In Section~\ref{sec:design}, the design
of the \pt\ source is described.  Details involving the fabrication of
the scandium tritide target and the assembly of the \pt\ source are
summarized in Section~\ref{sec:construction}.  The experimental setups
used in measuring the neutron and the gamma-ray output of the \pt\ source
are described in Section~\ref{sec:experimental}.  The results of
these measurements can be found in Section~\ref{sec:n_g_yield}, 
followed by the conclusions in Section~\ref{sec:conclusions}.

\section{Design Criteria for a High Energy Gamma-Ray Source}
\label{sec:design_criteria}

One way to calibrate the high-energy response (10$<E<$20~MeV) of a
large water \v{C}erenkov detector like SNO is to use high-energy gamma rays
generated from radiative-capture reactions induced by a particle beam.  The
devices that provide these high-energy gamma rays must be compact
enough to be maneuvered to different regions in the \hw\ volume using
the SNO calibration source manipulator system.  The largest insertion
port for calibration devices at SNO can accommodate devices up to
about 30~cm in diameter and 75~cm in length.  This physical constraint
limits the actual size of such calibration devices.

Because the SNO detector is essentially a 100\% efficient, 4$\pi$
detector to gamma rays in the solar neutrino energy regime, one does
not need to design a high-energy source with a high gamma-ray
production rate.  The centroid of the photopeak can be measured to
better than 1\% in less than an hour with a gamma-ray yield of
0.2~s$^{-1}$.

SNO is designed to run with MgCl$_{2}$ loaded in the heavy water to detect
the free neutron in Reaction (\ref{eq:nc}).  The high
energy gamma-ray source is required to have a low neutron production
rate.  This will minimize the signal interference of the gamma rays
resulting from thermal neutron capture by $^{35}$Cl in \hw\ in the
``salt'' running scenario and the dead time in the data acquisition
system.  A neutron production of less than 10$^{4}$~s$^{-1}$ is needed
for the design goal of $>$0.2~$\gamma$~s$^{-1}$.

The \pt\ source must be available to calibrate the SNO detector 
whenever there is a change to the detector configuration, or when a 
high energy calibration is called for.  An operational lifetime of 
$>$60~hours for the \pt\ source will be more than enough to calibrate 
the SNO detector during its anticipated life span.

Electromagnetic interference between this high energy calibration 
source and the photomultiplier tube array must be minimal.  For this 
reason, accelerator sources like the \pt\ source have to be run 
in direct current mode, instead of pulsed mode, to eliminate 
possible electromagnetic pickup by the photomultiplier tube array.

\section{Attributes of a $^3$H(p,$\gamma$)$^4$He Source}
\label{sec:attributes}

The $^3$H(p,$\gamma$)$^4$He reaction (see for example, Refs. 
~\cite{perry} and ~\cite{hahn}) has a Q-value of 19.8~MeV. Since 
$^4$He does not have a bound excited state, the gamma ray emitted in 
this reaction is monoenergetic.  Building a compact gamma-ray 
calibration source using this reaction is an attractive proposal for 
several reasons.

First of all, the projectile and the target have unit charge. 
Therefore, the effect of Coulomb suppression on the cross section for
this reaction is less than reactions with other combinations of
charged projectiles and targets.  Hence, the beam energy and power can
be minimised.  This allows the beam to be run in a d.c. mode without
incorporating a complicated cooling system for the target.

As the Q-value of $^3$H(p,n)$^3$He is -0.763 MeV, the \pt\ source is 
essentially neutron-free if the proton energy is below this 
threshold.  However, isotopic impurities and the isotopic exchange 
between the beam and the target will give rise to undesirable neutrons through 
the \dtn, \tdn, and \ttnn\ reactions.  In principle, one can eliminate 
this neutron production problem by mass analyzing the beam.  However, 
this option is not possible in the \pt\ source given the physical size 
constraint mentioned in the last section.

A monoenergetic calibration source like the \pt\ source is better than
sources with multiple energy lines in calibrating water \v{C}erenkov detectors
which generally have poor energy resolution.

\section{Design of the \pt\ Source}
\label{sec:design}

In order to keep the system as clean as possible, the \pt\ source was
built with ultra-high vacuum (UHV) hardware.  A cross sectional
drawing of the \pt\ source can be found in
Figure~\ref{fig:source_dwg}.  The source can essentially be divided
into three sections: the gas-discharge line, the ion acceleration line
and the target chamber.  In the following, the design of these three
sections is discussed.

The gas-discharge line is a cold-cathode Penning ion source, which
runs in d.c. mode with a very modest power consumption.  The outer
housing of the gas-discharge line consists of two
glass-to-stainless-steel adapters\footnote{Manufactured by Larson
Electronic Glass, Redwood City, CA, USA}.  Each of these adapters is
7.62~cm in length with a piece of 1.27-cm long Pyrex glass to isolate
the two ends.  The electrodes E1, E2 and E3 are welded to these
adapters.  The use of these glass-to-stainless-steel adapters provides
convenient high voltage isolation between the anode and the cathodes. 
The placement of the various electrodes in the gas-discharge line was
designed using the simulation program MacSimion~\cite{mcgilvery}.  In
the design, efforts were made to minimise ion loss to the electrode
walls; hence, a higher beam current can be attained for a given
discharge current.  The beam was spread over the target; this reduces
the areal power density and improves the target's longevity.  Under the normal
running scenario, the cathodes (E1 and E3) are kept at ground, whilst
the anode (E2) is maintained at +2~kV d.c.

A SAES St-172 getter (model LHI/4-7/200) is used as the hydrogen
discharge gas reservoir for the ion source.  The getter has 360~mg of
a zirconium-vanadium-iron alloy active material, and is mounted to
the BNC connector next to E1 in Figure~\ref{fig:source_dwg}.

The axial magnetic field required in the discharge is provided by a 
cylindrical magnet composed of seven 13.34~cm (outer diameter) by 
5.88~cm (inner diameter) by 1.91~cm (thick) barium ferrite Feroxdur 
ceramic rings\footnote{The magnets are supplied by Master Magnetics, 
Inc., Castle Rock, CO, USA. (part number CR525C)}.  The maximum magnetic field inside the central bore of 
the magnet is about 0.06~T. 

The ion acceleration line is a double-ended glass 
adapter\footnote{Manufactured by MDC Vacuum Products Corp., Hayward, 
CA, USA. (part number DEG-150).}, with 
one end attached to the gas-discharge line and the other connected to 
the target chamber which is biased at a negative high voltage.  In 
this scheme, the construction of complicated accelerating and focusing 
electrodes is avoided, and the length can also 
be kept to a minimum.  When the ions exit this acceleration line and 
enter the target chamber, they have acquired an energy equivalent 
to the target bias voltage, in addition to their ejection energy from the ion 
discharge region.

At the end of the ion acceleration line in the \pt\ source is the 
target mount flange.  The target is secured to a copper heat sink, as 
shown protruding from the flange in Figure~\ref{fig:source_dwg}, by a 
stainless steel screw-on cap.  This mounting mechanism is designed to 
allow efficient target mounting in the tritium glovebox in which this 
operation is to be performed.  

The total length of the \pt\ source is only 50~cm.  For deployment in 
SNO, it will be housed inside a 25.4-cm diameter by 60-cm long stainless 
steel cylindrical deployment capsule.  The dimensions of this capsule 
are well within the physical limits imposed by the SNO 
calibration-source-deployment hardware.

The expected yield of the \pt\ source was calculated.  Because the
cross section of the \ptg\ reaction below 50~keV is not well known,
the cross section at the operating voltage of the \pt\ source had to
be extrapolated from existing data.  It is shown in
Ref.~\cite{my_thesis} that the long-wavelength approximation formalism
developed in Christy and Duck~\cite{christy} is inadequate in
describing the \ptg\ cross section at low energies because of the
reaction's exceptionally high binding energy.  Using the lowest energy
data ($0.1\leq E_{p} \leq 0.75$~MeV) from Hahn~{\it et
al.}~\cite{hahn}, the reaction cross section $\sigma(E)$ was extracted
by performing a $\chi^{2}$ minimization of the S-factor $S(E)$, which is 
related to the cross section $\sigma(E)$ as~\cite{fowler}:
\begin{equation}
\label{eq:xsect_eqn}
      \sigma(E) \;=\; \frac{S(E)}{E}\exp\left( - \sqrt{\frac{E_G}{E}} \right),
\end{equation}
where $E$ is the energy in the center of mass frame, and $E_{G}$ is the Gamow energy.  Because $S(E)$ is expected to be a 
slowly varying function at low energy, it was fitted to the data as a power 
series:
\begin{equation}
\label{eq:s_factor}
S(E) \;\simeq\; S(0) \left( 1 + \frac{S^{\prime}(0)}{S(0)}E + \frac{1}{2}
            \frac{S^{\prime\prime}(0)}{S(0)}E^{2} \right ),
\end{equation}
where the parameters $S(0)$, $S^{\prime}(0)$ and $S^{\prime\prime}(0)$
were extracted.  Details of the extrapolation can be found in
Ref.~\cite{my_thesis}.  In Figure~\ref{fig:s_factors}, the
extrapolated $S(E)$ for the \ptg\ reaction is shown.  The values of
the fitted parameters are listed in Table~\ref{table:s_fitting}.  The
cross section at proton energies of 25~keV and 30~keV are 0.19~$\mu$b
and 0.30~$\mu$b respectively.  The stopping power required in the
yield calculation was calculated using the program SRIM~\cite{srim}. 
Figure~\ref{fig:estimated_pt_yield} shows the estimated gamma-ray
yield as a function of the the mass-1 content in a 50-$\mu$A, 27-keV
beam in the constructed \pt\ source.  This calculation assumed a total
mixing of hydrogen isotopes between the beam and the target.

The ion beam current was measured {\it in situ} by a calorimetric
method and by a Faraday cup fitted with a secondary electron
suppression scheme.  These measurements were made with extra hardware
installed in the target chamber of an untritiated model \pt\ source. 
In the calorimetric method, the temperature of a copper target, in
which a heater was embedded to calibrate the beam
power~\cite{my_thesis}, was monitored.  Beam current measured by both methods agreed
with each other.  The \pt\ source is capable of generating at least
50$\mu$A of total (atomic and molecular) beam current at a beam energy
of 20~keV. The mass composition of the beam was also measured {\it in
situ} by lengthening the target chamber and installing a home-built
mass spectrometer in the model source.  The mass-1 composition was
determined to be (0.63$\pm$0.09) in the H$_{2}$ partial pressure range
of 0.3$\times$10$^{-3}$ to 0.6$\times$10$^{-3}$~mbar, which is a
factor of $\sim$5 lower than the normal operating pressure of the \pt\
source.  The normal operating pressure of the source was chosen by
considering the beam stability and longevity running in a continuous
mode.  The mass composition measurement could not be made at the
normal operating H$_{2}$ pressure of the source due to increased beam
scattering in the lengthened target chamber and the inadequate
resolution of the spectrometer.

\section{Construction of the \pt\ Source}
\label{sec:construction}

\subsection{Fabrication of the Scandium Tritide Target}

The most common metal hydride films use titanium as the
``sorbent''~\cite{graves,jones,adair,yanagi,sumita}.  Singleton and
Yannopoulos~\cite{singleton} measured the loss rate of tritium in
titanium tritide, yttrium tritide and scandium tritide films at
elevated temperatures under several different ambient environments. 
It was demonstrated that both yttrium and scandium films have a lower
tritium loss rate than titanium films under the testing conditions. 
Although this study was performed using moderately loaded tritiated
films (Y:$^{3}$H and Sc:$^{3}$H ratios were $\sim$1:1), this general
observation of scandium tritide films having very good thermal
stability is believed to hold even for heavily loaded films.  This property is
essential for a target system which does not have an external cooling
mechanism like the \pt\ source.

Molybdenum was chosen as the substrate for the scandium film 
because of the strong adhesion between the two materials~\cite{frisch}.  
To ensure high adhesion strength of the scandium film to the 
molybdenum substrate, it was prepared by going through a series of 
mechanical and chemical treatments prior to film deposition.  

A substrate disc of diameter 2.86~cm was first cut out from a 1-mm 
thick sheet of 99.95\% pure molybdenum using the 
electro-discharge machining (EDM) technique.  This was to minimise the 
use of machining oil on the substrate.  The substrate was then 
sandblasted by fine glass beads in order to increase its effective 
surface area and enhance the film 
adhesion strength.  The scandium film would peel off much more easily 
from a non-roughened substrate surface.

The substrate was then treated chemically in a multi-stage process. 
It was first cleansed in acetone in an ultrasonic bath for half an
hour.  The substrate was subsequently ultrasonically cleansed in
ethanol, then deionised water, for half an hour in each solvent.  This
sequence of chemical cleansing ensured that hydrocarbons that might
have deposited on the substrate during the EDM process to be removed. 
The substrate surface was then etched in a 3~M nitric acid bath for
30~seconds.  The whole chemical cleansing process was completed by a
30-minute deionised water wash in an ultrasonic bath.

Once the substrate had gone through this series of preparation
processes, it was mounted to a copper holder in which a 110-W coil
heater was embedded and placed inside the ultra-high vacuum (UHV)
evaporation system which is described below.  The Mo substrate was
centered on the 2.54-cm diameter central aperture of the holder.  This
heater block was outfitted with thermocouples for temperature
monitoring.  The substrate was baked at 400$^{\circ}$C in the
evaporation system for about four days, then at 250$^{\circ}$C for
about a week to reduce outgassing from its surface.

Fabrication of the scandium tritide target, and the subsequent 
assembly of the \pt\ source were performed at the tritium laboratory 
at Ontario Hydro Technologies (OHT) in Toronto, Ontario, Canada.  The 
schematic of the vacuum system is shown in 
Figure~\ref{fig:oht_vacuum_system}.  To ensure that a high vacuum 
could be achieved in this tritium run, oil-free vacuum pumps 
and UHV hardware were used in this system.  The evaporation 
chamber is a UHV six-way cross with an outer flange diameter of 
15.24~cm.  The tritium-compatible glovebox is continuously purged 
with dry nitrogen.  The moisture level in the glovebox is typically 30 
to 50~ppm by volume.  The nitrogen purge gas is routed through a 
Zr$_{2}$Fe tritium trap in order to remove its tritium content before 
venting~\cite{tritium_workplan}.  The exhaust of 
the vacuum system is also routed through a Zr$_{2}$Fe trap before 
venting.

Two high-current feedthroughs were connected to the evaporation
chamber.  A 5-coil conical tungsten evaporation basket\footnote{R.D.
Mathis Company, Part Number B12B-3x.025W} was mounted between these
feedthroughs.  A (26$\pm$1)-mg lump of 99.99\% pure, sublimed
dendritic scandium was placed inside this basket, and positioned
directly above the molybdenum substrate in the heater block.  The
separation between the bottom of the tungsten basket and the
molybdenum substrate was (14$\pm$2)~mm.  A stainless steel shroud that
had an orifice directly below the evaporation basket was positioned
around the feedthrough-basket assembly to prevent deposition on the
viewport in the evaporation chamber and to reflect radiation back to
the coil to enhance heating efficiency.

A quartz oscillator was installed at the end of an evaporator bellows as shown in 
the setup in Figure~\ref{fig:oht_vacuum_system}.  When the deposition 
assembly is inserted into the evaporation chamber, the oscillator can 
be lowered to the back side of the assembly through an aperture in the 
main shroud and used to monitor the 
deposition rate of scandium.  The distance between the scandium source 
(in the tungsten evaporation basket) and the oscillator was 27~cm.  

As shown in Figure \ref{fig:oht_vacuum_system}, there are two main gas 
lines connected to the evaporation chamber in the vacuum system of the 
setup.  One of these branches is connected to a 5-g depleted uranium 
bed.  This uranium bed is used to store tritium which can be readily 
desorbed by raising it to sufficiently high 
temperature~\cite{ubed_1,ubed_2}.  In 
Table~\ref{table:tritium_gas_composition}, the isotopic purity of the 
tritium gas in this bed is shown.

Prior to film evaporation, the whole apparatus was baked for over a
week at $\sim$150-200~$^{\circ}$C to reduce the outgassing rate of the
evaporation system.  The tungsten evaporation coil was also baked by
running a 10~A current through it.  The base pressure of the system
was $\sim$6$\times$10$^{-7}$~mbar during the bakeout.  After the
baking, the evaporation system reached a base pressure of
5.8$\times$10$^{-8}$~mbar.

After bakeout, the deposition assembly (i.e. the high current 
feedthrough-evaporation basket assembly) was delivered into the 
evaporation chamber by winding in the linear translation stage to 
which the deposition assembly flange was connected.  The tungsten 
evaporation basket was positioned directly above the centre of the 
molybdenum substrate.  

The current fed to the tungsten basket was raised at a rate of about
1~A~min$^{-1}$ during the first thirty minutes of the experiment. 
This rate was then decreased to 0.2~A~min$^{-1}$ to lower the
outgassing rate of the evaporation hardware.  The basket
current was raised up to 46~A, at which point the coil temperature was
$\sim$1900~$^{\circ}$C. This was to ensure that all the scandium,
whose melting point is 1539~$^{\circ}$C, was evaporated. 

Immediately after the scandium deposition, the deposition assembly was
removed from the evaporation chamber by winding out the linear
translation stage and closing a gate valve (V2 in Figure
\ref{fig:oht_vacuum_system}).  Before tritium was let into the
evaporation chamber, the evaporation chamber was isolated by closing
the remaining gate valves (V1 and V3 in Figure
\ref{fig:oht_vacuum_system}) connected to it.  These two steps would
reduce the amount of tritium used in the subsequent tritiation
process.  The molybdenum substrate
temperature was subsequently raised to 400~$^{\circ}$C to enhance
tritium sorption by the scandium film later on.

The uranium tritide bed was first heated to 135~$^{\circ}$C to drive 
out the $^{3}$He from tritium decay in the bed.  At this temperature, 
tritium is still ``locked'' inside the bed.  The released $^{3}$He was 
first pumped out of the system before the uranium bed temperature was 
raised to 220-240~$^{\circ}$C at which temperature the tritium is 
desorbed.  In order to measure the amount of tritium sorbed by the 
scandium film, the tritium gas released from the uranium bed was first 
trapped in the small volume between valves V6 and V10 (see Figure 
\ref{fig:oht_vacuum_system}) before releasing to the isolated 
evaporation chamber.  This trap has a volume of 
(31.9$\pm$2.2)~cm$^{3}$.  With the tritium pressure measured by the 
pressure transducer connected to this volume, the amount of tritium 
used could then be determined.  In Figure \ref{fig:sct2_tritiation} the pressure inside the
evaporation chamber is plotted against the time after Doses 1, 7, 9
and 13 were injected.  It is clear from the figure that the sorbing
capacity of the scandium film decreased as the tritium concentration
in the film increased.  

A total of (8.19$\pm$0.57)~Ci of tritium gas was injected in 13
different doses into the chamber.  It was found that 89.9\% of the tritium
that was injected into the chamber was absorbed by the
(5.7$\pm$0.6)~mg scandium film on the target heater block.  This
corresponds to a $^{3}$H/Sc atomic ratio of (2.0$\pm$0.2).  The target
substrate subtended a smaller solid angle than the heater block to
which the substrate was mounted.  After correcting for the solid
angle, the tritium activity on the target substrate was found to be
(3.3$\pm$0.8)~Ci.

\subsection{Assembly of the \pt\ Source}

The ion source must be cleansed before it could accept the tritiated 
target.  If the 
outgassing rate of the ion source is too high, the getter would lose 
most of its capacity on pumping the residual gas in the source, 
rather than serving its purpose as the hydrogen discharge gas reservoir.

The ion source was cleansed chemically and mounted to a tritium-free
bakeout system.  The ion source was
baked at 150~$^{\circ}$C for about two weeks.  The bakeout vacuum
system was flushed with argon for approximately 5 to 10 minutes daily
during this bakeout period.  This flushing procedure did improve the
overall cleanliness of the vacuum system.

After the target fabrication, the ion source was removed from the 
bakeout system and wrapped in layers of Parafilm$^{\mbox{\tiny TM}}$ 
which is a flexible, thermoplastic material.  It was used to minimise 
tritiated particles depositing on the outer surface of the ion source 
once it was taken into the glovebox where the target evaporation 
system was set up.  The tritiated target was removed from the 
evaporation system and mounted to the \pt\ source.  The ion source was 
then connected to the vacuum system as indicated in Figure 
\ref{fig:oht_vacuum_system}.

After the system had reached its base pressure, H$_{2}$ was let into 
the system, and an ion beam was allowed to strike and to bombard the 
target for 5 minutes.  During this time, the beam energy was gradually 
increased from 0 to 25~keV. This procedure was necessary to cleanse 
the Penning electrodes by electro-discharge.  Contamination on the 
target, which might have deposited on the target surface during the 
target mounting process, would also be removed by this brief beam 
bombardment.  It was found that if this step were not carried out, the 
getter in the source would not be able to handle the residual gas load 
in the source once sealed.

The St-172 getter had to be activated before loading hydrogen to it. 
To activate the getter, it was heated for 10 minutes at 800$^{\circ}$C
by passing a 4.5~A current through it.  Once activated, the getter
current was lowered to about 1.6~A in order to maintain a temperature
of 200$^{\circ}$C. The getter was then loaded with hydrogen by
allowing an ambient H$_{2}$ pressure of 3.3$\times$10$^{-4}$~mbar into
the ion source.  After 30 minutes, $\sim$200~cm$^{3}$~mbar of H$_{2}$
would have been absorbed by the 360~mg of active material in the
getter.  The getter loading procedure was completed by turning off the
getter current, and by pumping out the residual H$_{2}$ gas in the ion
source.  After the base pressure was reached, the source was isolated
and detached from the rest of the vacuum system by closing the
metal-seal valve on the source.  The source was subsequently removed
from the glove box, and its outer surface was de-contaminated.

\section{Experimental Setup for Measuring the Neutron and Gamma-
Ray Yields of the \pt\ Source}
\label{sec:experimental}

\subsection{Gamma-Ray Detection Systems}

After the \pt\ source was constructed at OHT, a quality assurance test 
was first performed at Queen's University at Kingston, ON, Canada.  
The source was subsequently transported to the University of Washington for a 
measurement of the gamma-ray angular distribution in the \ptg\ 
reaction~\cite{my_thesis}.  In the quality assurance test, 
a 12.7-cm diameter by 7.6-cm long bismuth germanate 
(Bi$_{4}$Ge$_{3}$O$_{12}$, or BGO) crystal was used as the gamma-ray detector 
\cite{robk_thesis}.  In the angular distribution measurement,  
three 14.5-diameter by 17.5 cm cylindrical barium fluoride 
(BaF$_{2}$) crystals were used.  In Figures~\ref{fig:queen_bgo_scheme} 
and~\ref{fig:baf2_arrangement} the orientations of the \pt\ source 
with respect to the detectors in the two different test systems are 
shown.  

\subsection{Neutron Detection System}

Because of beam-target mixing, fast neutrons are generated 
through the \dtn, \tdn, and \ttnn\ reactions.  The 
neutron output of the \pt\ source during its lifetime was monitored by 
neutron-proton elastic scattering in organic scintillators.  The 
neutron detector was a 12.7-cm diameter by 5.1-cm thick Bicron BC~501 
liquid scintillator, which was optically coupled to a Hamamatsu R1250 
photomultiplier tube (PMT).

A Piel~112 pulse shape discriminator (PSD)~\cite{piel} was used to
perform pulse shape discrimination on gammas and fast neutrons
generated by the \pt\ source.  The neutron-gamma separation ability in
the neutron detection system is demonstrated in
Figure~\ref{fig:n_gamma_sep_tac}.

\section{Gamma-Ray and Neutron Yields of the Source}
\label{sec:n_g_yield}

The gamma-ray and neutron production rates by the \pt\ source are
summarised in this section.  The \pt\ source was operated in the
quality assurance test and in a measurement of the gamma-ray angular
distribution in the \pt\ reaction.  During the 98.8~hours of
operational lifetime of the \pt\ source, data was taken at beam energies of 22,
27 and 29~keV.

\subsection{Gamma-Ray Yields}
\label{sec:g_yield}

In the quality assurance test at Queen's, the source was run at a beam
energy of 22~keV for 3 hours.  The gamma-ray output was subsequently
increased by raising the beam energy to 27~keV, and ran for another
17.9~hours.  Energy calibration of the BGO detector was provided by
the $^{22}$Na 0.511-MeV and 1.275-MeV lines, the
$^{1}$H(n,$\gamma$)$^{2}$H 2.22-MeV line, and the
$^{12}$C$^{\ast}$(4.4~MeV) de-excitation line.  In
Figure~\ref{fig:queens_bgo_mc_fit} the 
cosmic-ray-background-subtracted energy spectrum from part of
the data taken at 27~keV in the quality assurance test is shown.  The figure shows a fit using a
response function for the BGO spectrometer generated by
GEANT~\cite{geant_manual}.  The measured gamma-ray yield of the \pt\
source during its testing at 27~keV is (0.67$\pm$0.11)~s$^{-1}$.  The
gamma-ray yield at 22~keV could not be extracted because of low
statistics.

In the angular distribution measurement, the gamma-ray detectors were energy calibrated by a variety of sealed
sources: $^{137}$Cs(0.662~MeV), $^{207}$Bi(1.063~MeV),
$^{12}$C$^{\ast}$(4.44~MeV), and $^{16}$O$^{\ast}$(6.13~MeV).  Without
a readily available energy source with an energy close to 19.8~MeV,
Monte Carlo simulation using GEANT~\cite{geant_manual} was relied upon
to calculate the response of the detectors.  The simulation program was
checked against the data taken with a strength calibrated
$^{13}$C($\alpha$,n)$^{16}$O$^{\ast}$ source.

Energy spectra were taken with this
$^{13}$C($\alpha$,n)$^{16}$O$^{\ast}$ source placed at the centre of
the \baftwo\ detector system.  At the time of this experiment, this
source had a strength of
(4.1$\pm$0.1)$\times$10$^{3}$~$\gamma$~s$^{-1}$.  Because of its high
neutron output, energy spectra were taken with a 2.5-cm thick slab of
lead placed between the source and the detectors to extract the
neutron induced spectra.  By comparing these two types of spectra, the
gamma-ray line shape could then be extracted for each detector.  In
Figure \ref{fig:mc_line_shape_compare}, the GEANT generated line shape
is compared to an experimentally determined spectrum.  After
correcting for the effects of lead absorption, neutron induced
background and dead-time, the number of detected gamma rays and
efficiency ($\varepsilon_{exp}$) were extracted.  The average ratio
between $\varepsilon_{exp}$ and the GEANT calculated efficiency
($\varepsilon_{MC}$), $\varepsilon_{exp}/\varepsilon_{MC}$, was found
to be (1.01$\pm$0.04).

The gamma-ray penetration function $\eta_{\gamma}(\theta)$ was measured for the
6.13-MeV gamma-ray line in the three \baftwo\ detectors.  This source
was positioned inside an untritiated model \pt\ source, the
mechanical construction of which was identical to the real \pt\ source, at the
location where the tritiated target would be mounted.  The gamma-ray
detection rate was then measured experimentally in a procedure similar
to the efficiency measurement above.  By comparing this detection rate
and the one without the presence of the model source, the average
penetration factor over the solid angle subtended by the detectors
$\langle \eta_{\gamma}(\theta) \rangle_{\Omega_{det}}$ was 
extracted.  The average percentage difference between the measured
values and the simulated ones is $\sim\pm$3\%.

To extract the gamma-ray yield of the \pt\ source, the 
calibrated ``beam-on'' data were fitted to a composition of a cosmic-ray 
background and the 19.8-MeV line shape for an isotropic source 
located at the target surface in the \pt\ source as generated by GEANT 
simulation.  Because the emitted gamma rays in the \ptg\ reaction have 
a predominant $\sin^{2}\theta$ angular 
distribution~\cite{perry,my_thesis,delbianco}, the extracted 
gamma-ray amplitude from the fit was corrected for this distribution.

The rate at $E_{p}=$29~keV during the gamma-ray angular distribution
measurement in the last 47.2 hours of the source's lifetime was
(0.36$\pm$0.03)~s$^{-1}$.  The gamma-ray production rate in between
the quality assurance run and this time was not evaluated because of a
noise problem in the electronics system, and the yield could not be
extracted reliably.  In Figure~\ref{fig:neutron_generation_scaled},
the gamma-ray production rate was renormalised to that for a 29-keV
atomic beam.  It is clear that the gamma-ray yield decreased over time
and is due to beam-target mixing and target sputtering in the source. 
This point will be discussed after evaluating the neutron yields in
the next section.

\subsection{Neutron Yields}
\label{sec:n_yield}

In the \pt\ source most of the neutrons are generated through the
$^{3}$H+$^{3}$H interaction.  Although the discharge gas stored into
the hydrogen reservoir in the \pt\ source was initially free of any
tritium, tritium would get into the discharge gas through beam-target
exchange after a period of beam bombardment.  Moreover, deuterium
present in the discharge gas (at a 1.5$\times$10$^{-4}$ level) and in
the target (at a 1.2$\times$10$^{-3}$ level) would enhance neutron
production by the source through the \tdn\ reaction.  In the following the results of
this neutron production measurement are presented.

The fast neutron detection efficiency of the liquid scintillator was 
calibrated using an $^{241}$Am-$^{9}$Be source which generates 
neutrons through $^{9}$Be($\alpha$,n)$^{12}$C. This source has 
a calibrated neutron strength of 
(7.1$\pm$0.7)$\times$10$^{3}$~n~s$^{-1}$ and was placed on 
the axis of the detector with a separation of 20.6~cm, the 
same distance between the tritiated target and the neutron detector in 
the gamma-ray angular distribution runs.  Gamma rays and neutrons 
generated by the source could be cleanly separated by pulse shape 
discrimination.  The net 
neutron count rate was extracted after the correction of a 
(7.1$\pm$0.1)\% dead time and the subtraction of a background rate of 
0.7~s$^{-1}$.  The detection efficiency 
($\varepsilon\Delta\Omega/4\pi$) was found to be $(3.6 \pm 0.4) \times 
10^{-3}$.

Neutrons generated by the \pt\ source would inevitably be scattered or
absorbed by its construction material.  Hence the detected neutron
rate ($R_{det}$) would be less than the actual \pt-source generated rate
($R_{gen}$) by a reduction factor $\eta_{n}$.  To measure this
reduction coefficient, the $^{241}$Am-$^{9}$Be source was placed on
the target mount inside the untritiated model source.  This model
source was then placed in the same orientation to the liquid
scintillator as in the gamma-ray angular distribution runs.  After
correcting for the dead time and background, and comparing the neutron
detection rate to that in the calibration runs without the presence of
this model source, it was found that the \pt\ source hardware absorbed or
scattered $(38 \pm 6)$~\% of the neutrons that are generated inside
the source.

Because there was a variation in beam intensity on target from run to
run, the neutron production rate was normalised to the current drawn
from the target bias supply in order to provide a fair comparison. 
This current was a combination of the actual ion current on target and
the contribution from secondary electron emission.  This current was
monitored during all the experimental runs.  The \pt\ source does not
have any internal secondary electron suppression scheme because of
physical constraints imposed by the SNO calibration hardware.

Two assumptions were made in extracting this neutron generation rate
by the \pt\ source:
\begin{enumerate}
	\item the neutrons generated by the \pt\ source have the same energy 
	spectrum as fast neutron spectrum from the $^{241}$Am-$^{9}$Be 
	calibration source;
	\item the angular distribution of neutrons generated by the \pt\ 
	source is isotropic as in the $^{241}$Am-$^{9}$Be case.
\end{enumerate}

Neutrons are produced 
predominantly by the $^{3}$H+$^{3}$H interaction in the \pt\ source.  
The reactions that are energetically possible in this system are:
\begin{eqnarray}
	\label{eqn:ttnn}
	&& ^{3}\mbox{H}(t,nn)^{4}\mbox{He} \\
	\label{eqn:ttnone}
	&& ^{3}\mbox{H}(t,n_{1})^{5}\mbox{He}^{\ast}(n)^{4}\mbox{He} \\
	\label{eqn:ttnzero}
	&& ^{3}\mbox{H}(t,n_{0})^{5}\mbox{He}(n)^{4}\mbox{He}.
\end{eqnarray}

In a measurement at a triton energy $E_{t}$=500~keV, the branching
ratio for these reactions was found to be 70\%:20\%:10\% (in the same
order as they appear above)~\cite{wong}.  The neutron energy spectrum
for each of these reactions is somewhat different.  Without any
final-state effect, the direct three-body breakup reaction in reaction
(\ref{eqn:ttnn}) would yield neutrons at an average energy of
$\sim\frac{1}{2}\cdot\frac{5}{6}Q$.  With a $Q$-value of 11.3~MeV, the
neutron energy spectrum from reaction (\ref{eqn:ttnn}) would be a
broad peak centered at about 4.7~MeV. This shape is indeed very
similar to the neutron spectrum from $^{9}$Be($\alpha$,n)
sources~\cite{anderson}.

The ground state transition (\ref{eqn:ttnzero}) yields a 10.4-MeV
neutron $n_{0}$, followed by a 0.9-MeV secondary neutron.  The neutron
detection efficiency for the liquid scintillator is almost null at 0.9
MeV. Reaction (\ref{eqn:ttnone}) is a sequential decay proceeding
through a broad $^{5}$He excited state at about 2~MeV. Because of the
small branching ratio for this excited state transition, it would not
contribute much to the uncertainty in the extracted neutron generation
rate by the \pt\ source.  The uncertainty in the extracted \pt-source neutron
rate due to the secondary neutrons is at most 15\% if one assumes none
of the secondary neutrons from (\ref{eqn:ttnone}) and
(\ref{eqn:ttnzero}) were detected.  The uncertainty in the extracted
\pt-source neutron rate due to $n_{0}$ and the 14~MeV monoenergetic
neutron from \tdn\ was estimated to be 9\% at $E_{p}$=29~keV.

Although the neutron detector was placed in different orientations to the \pt\ 
source in the quality assurance runs and in the 
gamma-ray angular distribution measurement, continuous beam-target 
exchange rendered it impossible to extract the neutron angular 
distribution without the presence of a second neutron detector for 
normalisation purposes.  Wong {\it et al.}~\cite{wong} measured the 
angular distribution for the  $^{3}$H+$^{3}$H system at 
$E_{t}$=500~keV.  They found that the ground state transition neutron 
group is isotropic to within an accuracy of $\pm$10\%.  
They also found that in the neutron energy range of 2 to 7.5~MeV, 
the continuum neutron group is also isotropic to within an accuracy 
of $\pm$20\% in the laboratory angle range of 4 to 100$^{\circ}$.  For 
the \tdn\ reaction, the angular distribution is isotropic at and below 
the resonance \cite{lauritsen}.  Given these facts, the 
assumption that the neutrons emitted by the \pt\ source are isotropic 
was made in the yield evaluation.

In order to look at the time variation of the neutron production rate
by the \pt\ source more closely, the neutron production rates for all
the runs were renormalised to the same atomic beam energy at 29~keV.
In other words, the rate in all of the $E_{p}$=22~keV and 27~keV runs
were scaled up by a factor corresponding to the difference in cross
section at that atomic beam energy to that at 29~keV. The resulting
plot is shown in Figure~\ref{fig:neutron_generation_scaled}.

In Figure~\ref{fig:neutron_generation_scaled}, it is clear how the
neutron production rate in the \pt\ source varied over time.  The
neutron production rate was gradually increasing initially.  This is a
clear indication of beam-target exchange, as tritium in the target
gets into the discharge gas stream.  The neutron production rate then
began to decrease.  This can be explained by the fact that the rate of
hydrogen isotope exchange was reaching an equilibrium, and sputtering
of the target became the dominant process.  The target sputtering
effect had caused the build-up of a thin film on the high voltage
insulator in the acceleration section of the source.  Under the normal
operating condition, one end of this insulator is grounded whilst the
other end is biased at $\sim$-30~kV.  With a thin conductive film
build-up on the insulator, high voltage could no longer be maintained
without breakdown.  This build-up effect limited the lifetime of the
\pt\ source to 98.8~hours\label{pg:sputter_target}.

The neutron production rate of the \pt\ source during calibration in
the SNO detector was estimated.  Using the highest data point in
Figure \ref{fig:neutron_generation_scaled}, the maximum neutron
generation rate was estimated to be less than
(2.5$\pm$0.4)$\times$10$^{3}$~n~s$^{-1}$.  The uncertainty here does
not include the monoenergetic neutron and the secondary neutron
contributions discussed above.  However, the estimated rate quoted
above should be seen as the upper limit of neutron production as it
was estimated using the highest data point in the data.  In
Figure~\ref{fig:pt_spec_total}, the results of a Monte Carlo
simulation of the SNO detector response to neutrons and gamma rays
generated by the \pt\ source are shown.  This simulation was performed
using the SNO Monte Carlo and analysis program SNOMAN~\cite{snoman}. 
In this simulation, fast neutrons generated by the \pt\ source were
assumed to be monoenergetic at 4.7~MeV. Full \pt\ source and
deployment capsule geometries were employed in this simulation, but
neutron absorbers inside the source's stainless steel deployment
housing were not.  This is equivalent to assuming the worst possible
neutron leakage into the heavy water.  A neutron production rate of
2,500~s$^{-1}$ and a gamma-ray production rate of 0.6~s$^{-1}$ were
assumed.  The spectra in the figure represent about 3~hours of run
time in the SNO detector.  From these figures, it is clear that the
neutron production rate of the \pt\ source is low enough for an 
accurate measurement of the 19.8-MeV photopeak.

\section{Conclusions}
\label{sec:conclusions}

A functional 19.8-MeV gamma-ray source using the \ptg\ 
reaction was built.  This \pt\ source met all the physical 
and operational requirements for energy calibration at the SNO detector.   
This is the first self-contained, compact and portable high energy 
($E_{\gamma}>$10~MeV) gamma-ray source of this type.  

Techniques to fabricate high-quality 
scandium deuteride and tritide targets were developed.  The tritiated 
target had a Sc:$^{3}$H atomic ratio of 
1:2.0$\pm$0.2.  

In the testing of the \pt\ source, 19.8-MeV gamma rays from the \pt\
reaction were observed and found to be sufficient for calibrating the
SNO detector.  The neutron production rate by the \pt\ source is also
low enough that the neutron background would not mask the gamma-ray
signal during calibration.  Because of the time variation of its
output, this \pt\ source is not suitable for efficiency calibration.

The operational lifetime of the \pt\ source was 98.8~hours.  Operation
was terminated by a thin conducting layer deposited on the high
voltage insulator in the ion acceleration line, which caused a high
voltage breakdown across the insulator.  The origin of this layer was
scandium sputtering off the target surface.  A second \pt\ source has
been constructed with minor engineering changes to reduce this
deposition effect.  

Calibration of large water \v{C}erenkov detectors at 
energies near the solar neutrino endpoint has been a 
difficult problem.  This proof-of-principle experiment of the 
\pt\ source opens a window for more convenient calibration 
standards in the future.  One area in which the \pt\ source can be 
improved is to implement a beam analyser to reduce the beam 
power on the target, and to reduce the neutron output of the \pt\ 
source.  This feature was not instrumented in this project because of 
stringent constraints on the physical size of calibration sources 
that can be deployed in the SNO detector.

\begin{ack}
 
We thank Mel Anaya, Tom 
Burritt, Mark Hooper, Clive Morton, Hank Simons, and Doug Will for 
their technical support at various stage of this project.  We thank 
David Sinclair for his careful reading of the manuscipt and his valuable 
comments.  One of us (AWPP) would like to thank the University of British 
Columbia for a University Graduate Fellowship.  This work was 
supported by the Natural Sciences and Engineering Research Council of 
Canada, and by the US Department of Energy under Grant Number DE-FG06-90ER40537.   
    
\end{ack}

\clearpage


\begin{table}
\begin{center}
\caption[$\chi^{2}$ minimisation results in fitting the S-factors from 
Hahn~{\it et al.}]{$\chi^{2}$ minimisation results in fitting the 
S-factors from Hahn~{\it et al.}  }
\protect\label{table:s_fitting}
\begin{tabular}{lccc} \hline\hline
 & $S(0)$ & $S^{\prime}(0)/S(0)$ & 
 $\frac{1}{2}S^{\prime\prime}(0)/S(0)$ \\ 
 &  (MeV b) & (MeV$^{-1}$) & (MeV$^{-2}$) \\ \hline
Fitted values & (1.30$\pm$0.40)$\times$10$^{-6}$ & 
                  25$\pm$12 & 38.8$\pm$7.3 \\ \hline 
Correlation &  1.00 & -0.99 & -0.65 \\
Matrix     &  -0.99 & 1.00 & 0.53\\ 
           &  -0.65 & 0.53 & 1.00 \\ \hline
$\chi_{\nu}^{2}$ & \multicolumn{3}{c}{0.22} \\ \hline\hline
\end{tabular}
\end{center}
\end{table}

\clearpage

\begin{table}
\begin{center}
\caption{Isotopic composition of the tritium gas used in the target.}
\protect\label{table:tritium_gas_composition}
\begin{tabular}{|c|l|} \hline
Isotope & Composition \\ \hline
$^{1}$H & (0.79$\pm$0.04)\% \\
$^{2}$H & (0.12$\pm$0.01)\% \\
$^{3}$H & (99.09$\pm$0.05)\% \\ \hline
\end{tabular}
\end{center}
\end{table}

\clearpage

\begin{figure}
    \begin{center}
    \epsfxsize=5.25in
	\epsfbox{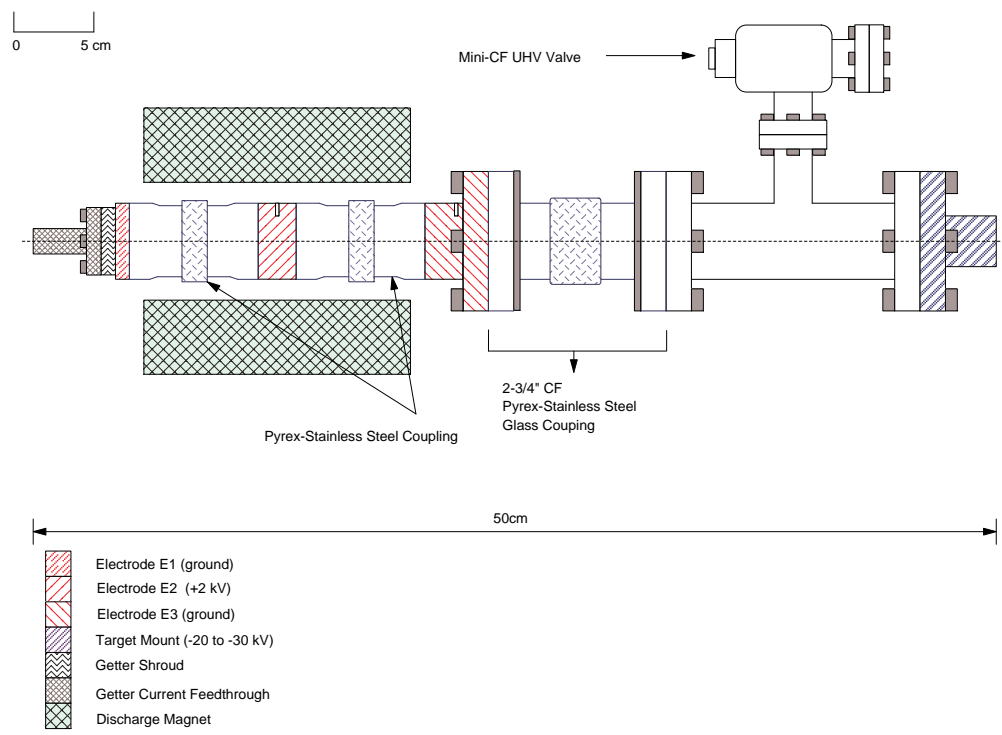} 
	\caption{Cross sectional drawing of the \pt\ source.}
	\protect\label{fig:source_dwg}
	\end{center}
\end{figure}
\clearpage

\begin{figure}
    \begin{center}
	\epsfysize=3.25in 
	\epsfbox{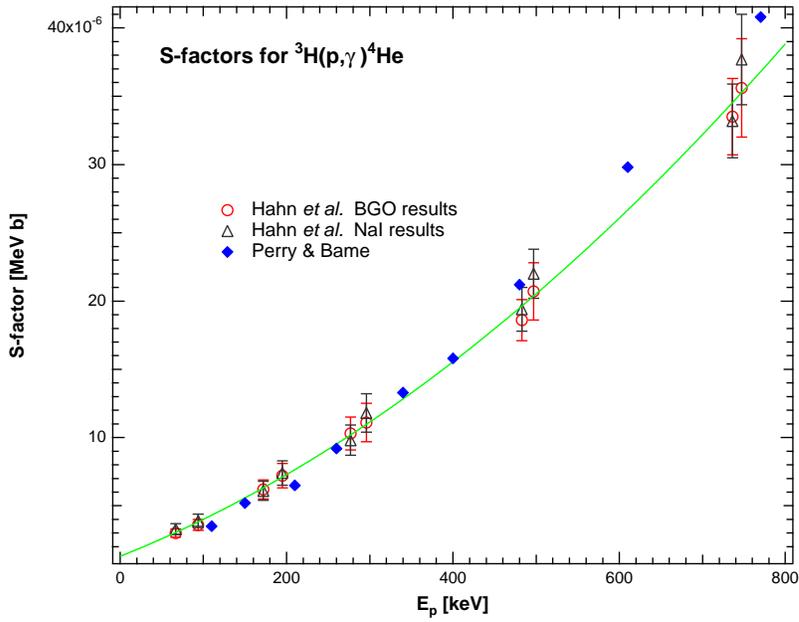} 
	\caption[Extrapolated S-factors for the \ptg\ reaction]
	{Extrapolated S-factors for the \ptg\ reaction.  Measured
	S-factors by Perry and Bame~\cite{perry} and Hahn~{\it et
	al.}~\cite{hahn} are shown as data points in this plot. 
	Hahn~{\it et al.} used a BGO detector and a NaI detector in
	their measurements, and the results for these two detectors
	are shown separately here.  The solid curve is the $\chi^{2}$
	fitted curve to the combined data in Hahn~{\it et al.} }
	\protect\label{fig:s_factors}
	\end{center}
\end{figure}

\clearpage

\begin{figure}
    \begin{center}
	\epsfysize=3.25in 
	\epsfbox{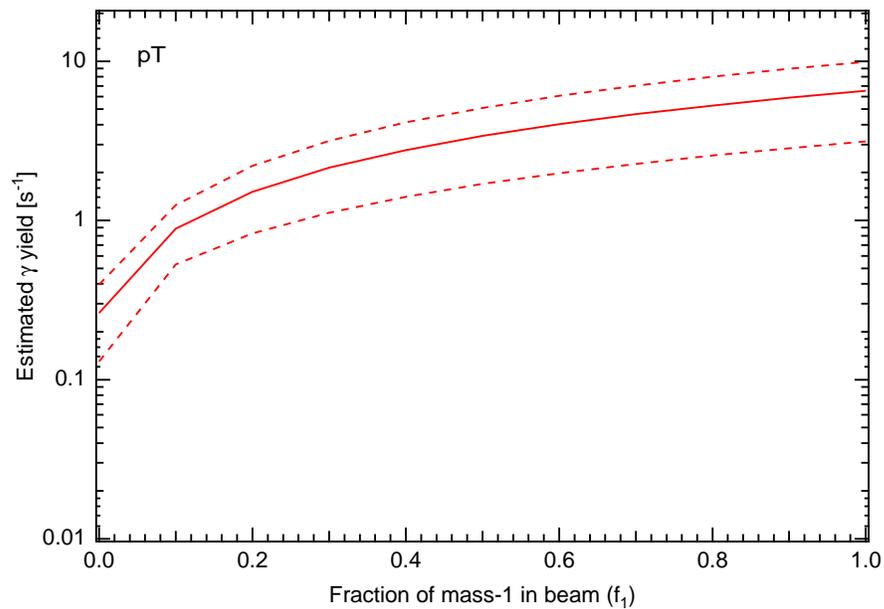} 
	\caption[Estimated gamma-ray yield from the \pt\ source]{Estimated 
	gamma-ray yield from the \pt\ source.  The yield is plotted 
	against the mass-1 fraction $f_{1}$ in a 50$\mu$A, 27-keV beam.  Hydrogen 
	isotopes in the beam and the target were assumed to be completely 
	mixed.  The yield shown here should be treated as the upper limit 
	because target degradation was not taken into account in 
	the calculation.  The dotted lines are the calculated 
	uncertainties based on the uncertainties in the physical 
	parameters of the constructed \pt\ source and the cross 
	section (\cite{my_thesis}).}
	\protect\label{fig:estimated_pt_yield}
	\end{center}
\end{figure}

\clearpage

\begin{figure}
    \begin{center}
    \epsfysize=5.5in
	\epsfbox{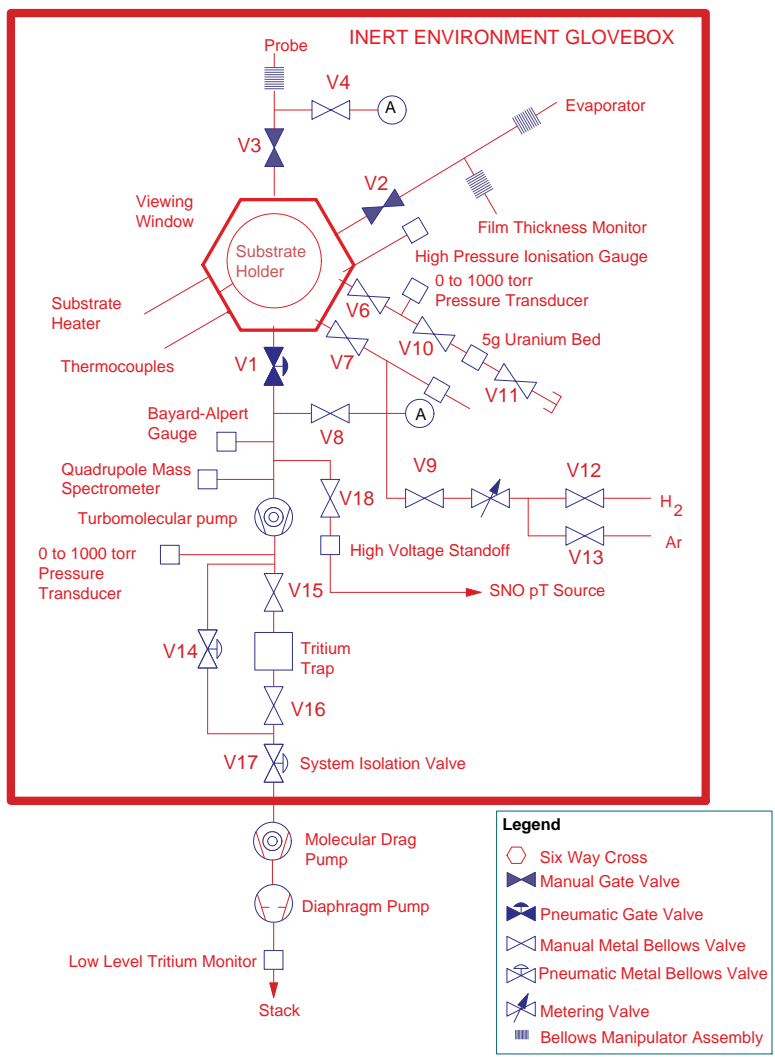} 
	\caption{Schematic of the scandium tritide target evaporation 
	vacuum system.  Most of the setup is enclosed in a dry nitrogen 
	environment inside a glovebox (from~\cite{tritium_workplan}).}
	\protect\label{fig:oht_vacuum_system}
	\end{center}
\end{figure}

\clearpage

\begin{figure}
    \begin{center}
	\epsfxsize=4.5in \epsfbox{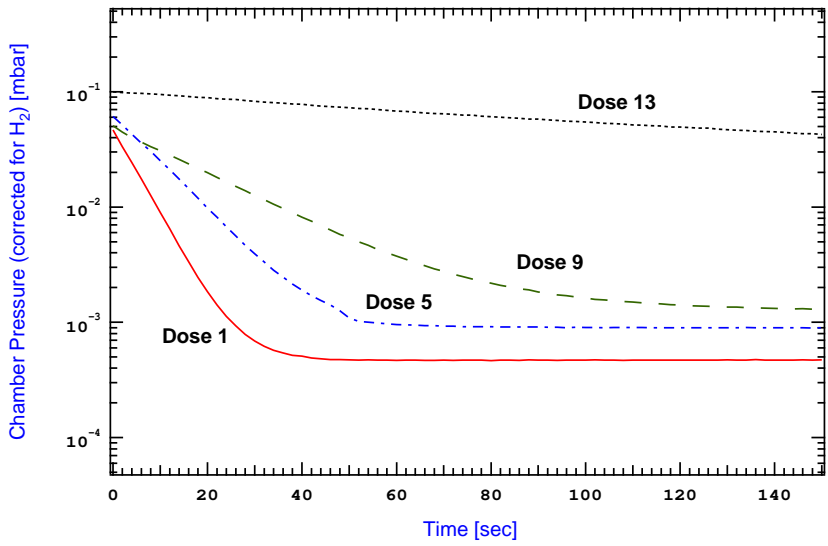}
	\caption[Tritium pumping by the scandium film]{Tritium pumping
	by the scandium film.  The pumping curve for Doses 1, 5, 9 and
	13 are shown here.}
	\protect\label{fig:sct2_tritiation}
	\end{center}
\end{figure}

\clearpage

\begin{figure} 
\begin{center} 
\epsfxsize=4.75in 
\epsfbox{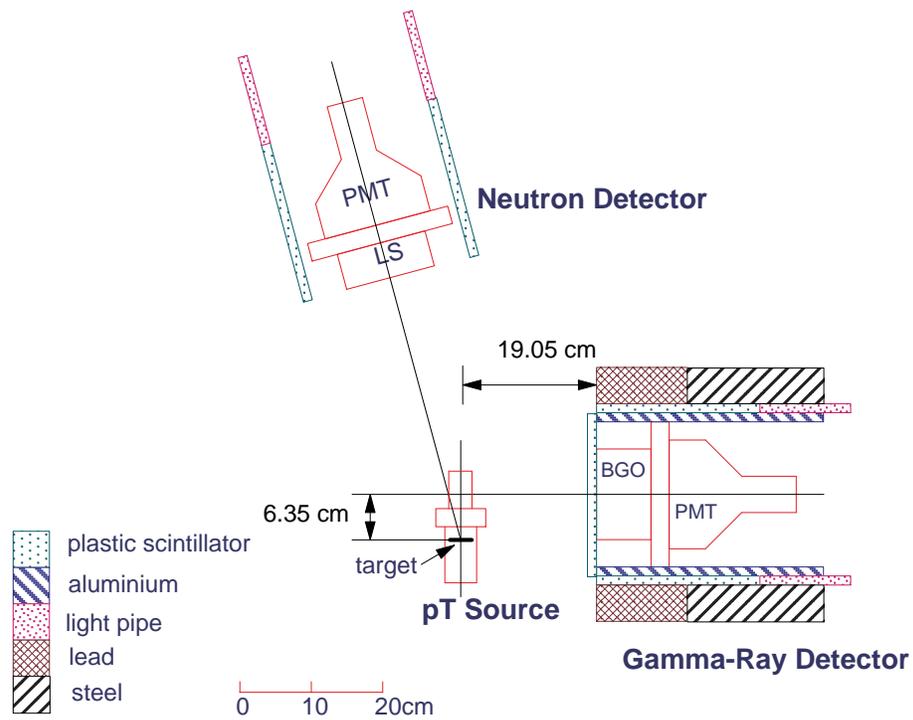} 
\caption{Top 
view of the BGO detector setup for the quality assurance testing of the \pt\ 
source.  The separation between the liquid scintillator (LS) and the target 
of the \pt\ source is about 36~cm.} 
\protect\label{fig:queen_bgo_scheme} 
\end{center} 
\end{figure} 

\clearpage

\begin{figure}
    \begin{center}
	\epsfysize=4in 
	\epsfbox{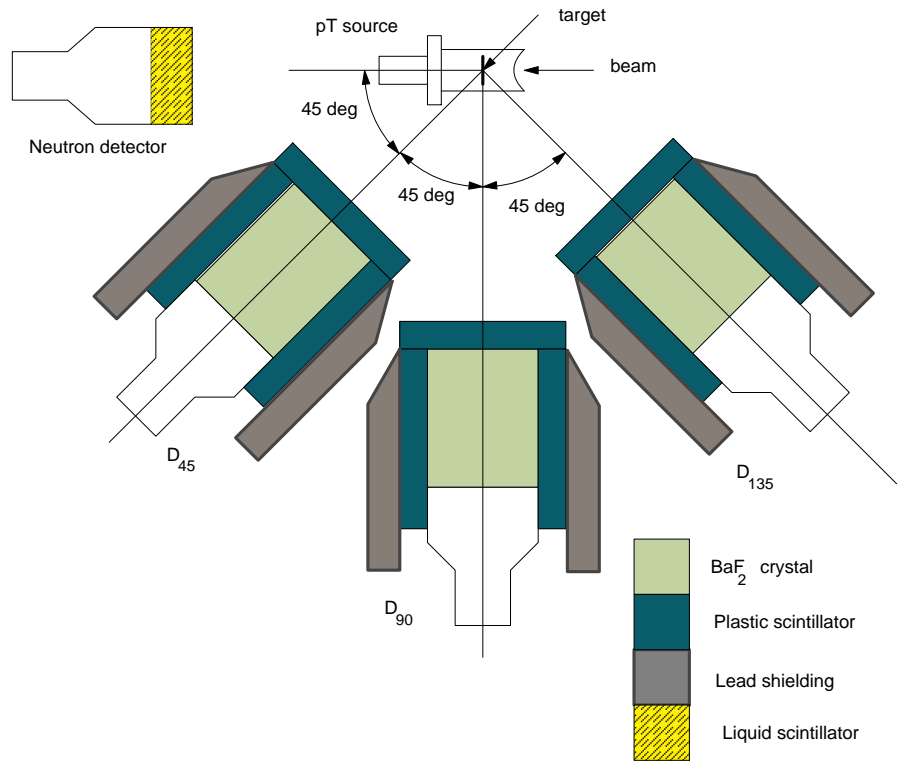} 
	\caption{Schematic 
	of the BaF$_{2}$ detector system.  The three BaF$_{2}$ detectors 
	were oriented at 45$^{\circ}$ (\detthree), 90$^{\circ}$ (\dettwo), 
	and 135$^{\circ}$ (\detone) to the beam direction, whilst the 
	neutron detector was oriented at 2$^{\circ}$ to the beam 
	direction.  The separation between the centre of the target and 
	the front face of the BaF$_{2}$ crystals was 35.6~cm for \dettwo, 
	and 25.4~cm for \detthree\ and \detone.  The neutron detector was 
	located at 20.6~cm from the centre of the target.}
	\protect\label{fig:baf2_arrangement}
	\end{center}
\end{figure}

\clearpage

\begin{figure}
    \begin{center}
    \epsfysize=3.25in
	\epsfbox{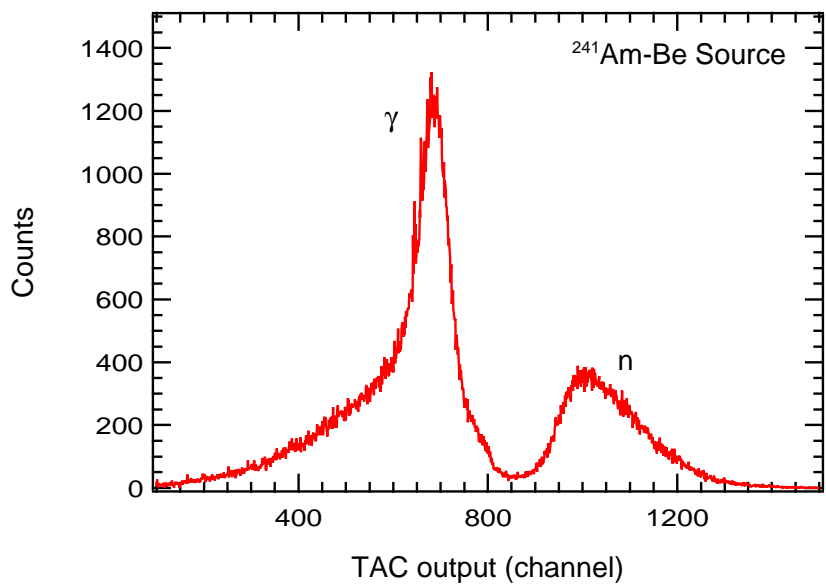} 
	\caption[Timing distribution of liquid scintillator pulses generated 
	by neutrons and gamma rays in a $^{9}$Be($\alpha$,n)$^{12}$C$^{\ast}$ 
	source]{Timing distribution of liquid scintillator pulses generated 
	by neutrons and gamma rays in a $^{9}$Be($\alpha$,n)$^{12}$C$^{\ast}$ 
	source.  Neutrons are cleanly separated from the gamma-rays using the 
	pulse shape discrimination scheme outlined in the text.}
	\protect\label{fig:n_gamma_sep_tac}
	\end{center}
\end{figure}

\clearpage

\begin{figure}
    \begin{center}
	\epsfysize=3.75in 
	\epsfbox{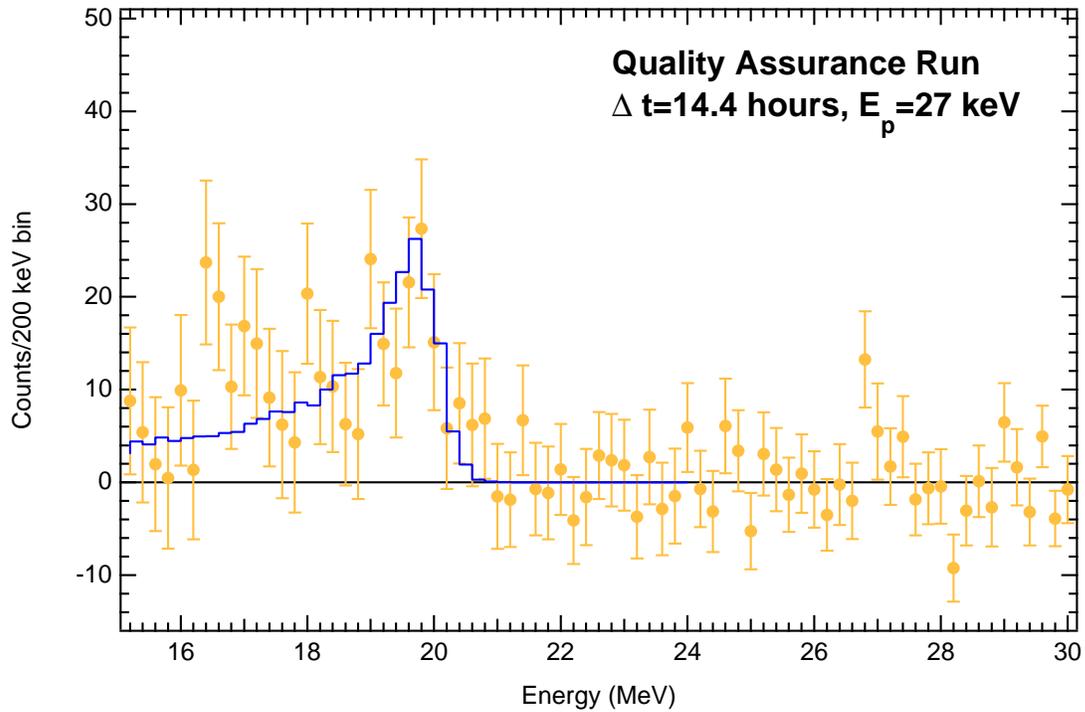}
	\caption[Background-subtracted BGO energy spectrum in the
	quality assurance run at Queen's
	University]{Background-subtracted BGO energy spectrum in the
	quality assurance run at Queen's University.  The data points
	constitute the background-subtracted energy spectrum.  The
	histogram shown is a fit using a response function for the BGO
	spectrometer generated by GEANT. The measured yield of the
	\pt\ source during its running at 27~keV is
	(0.67$\pm$0.11)~s$^{-1}$.  The excess near 16~MeV was due to
	statistical fluctuation, as this was not observed in the later
	running of the source.}
	\protect\label{fig:queens_bgo_mc_fit}
	\end{center}
\end{figure}

\clearpage

\begin{figure}
    \begin{center}
	\vskip -0.25in \epsfysize=3in 
	\epsfbox{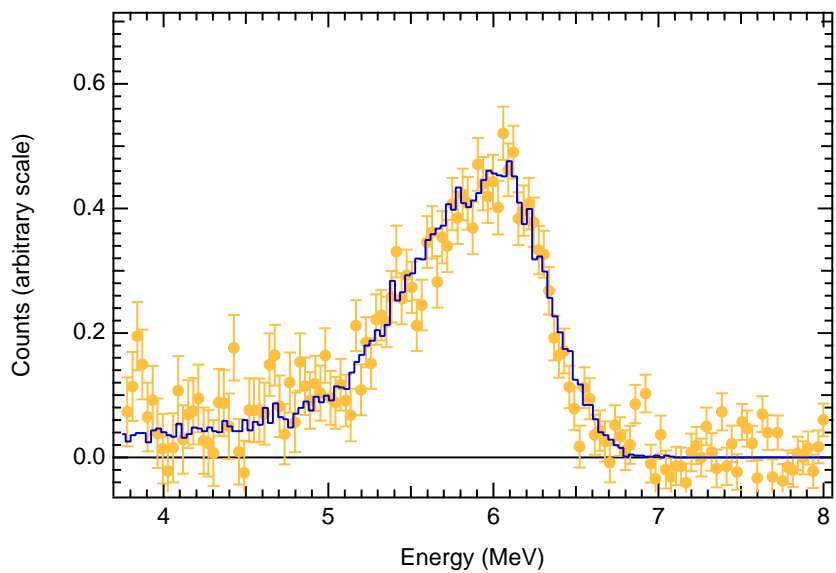} \caption[Comparing 
	GEANT generated gamma-ray line shape to measurement]{Comparing 
	GEANT generated gamma-ray line shape to measurement.  The data 
	points correspond to the 6.13-MeV line from a calibrated 
	$^{16}$O$^{\ast}$ de-excitation source.  The solid histogram is 
	the GEANT generated line shape.}
	\protect\label{fig:mc_line_shape_compare}
	\end{center}
\end{figure}

\clearpage

\begin{figure}
    \begin{center}
	\epsfxsize=4.25in \epsfbox{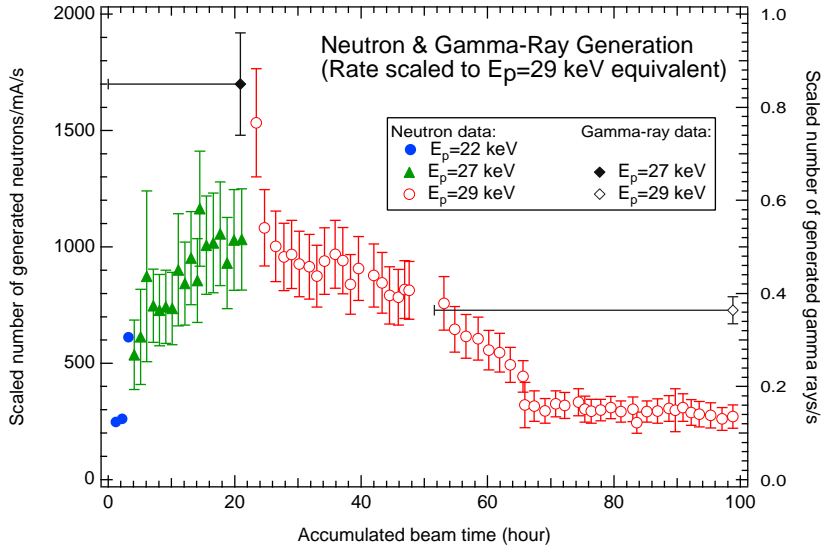}
	\caption[Scaled neutron and gamma-ray production by the \pt\
	source at $E_{p}$=29~keV]{Scaled neutron and gamma-ray
	production by the \pt\ source at $E_{p}$=29~keV. The rates
	were normalised to the current drawn from the target power
	supply during the runs.  Also, the production rates for the
	$E_{p}$=22~keV and 27~keV runs have been scaled to the
	$E_{p}$=29~keV level.  The scaling was done by assuming a pure
	atomic beam of protons or tritons since the contribution to
	the signals from molecular ions are much smaller.  The ``error
	bars'' on the accumulated beam time for the gamma-ray results
	represent the time intervals in which the mean production
	rates were calculated.  The gamma-ray yield could not be extracted 
	reliably between the ~20-th and the 50-th hour of the source lifetime 
	because a noise problem in the electronics system, which was 
	subsequently eliminated.}
	\protect\label{fig:neutron_generation_scaled}  
	\end{center}
\end{figure}

\clearpage

\begin{figure}
    \begin{center}
	\epsfxsize=5.in 
	\epsfbox{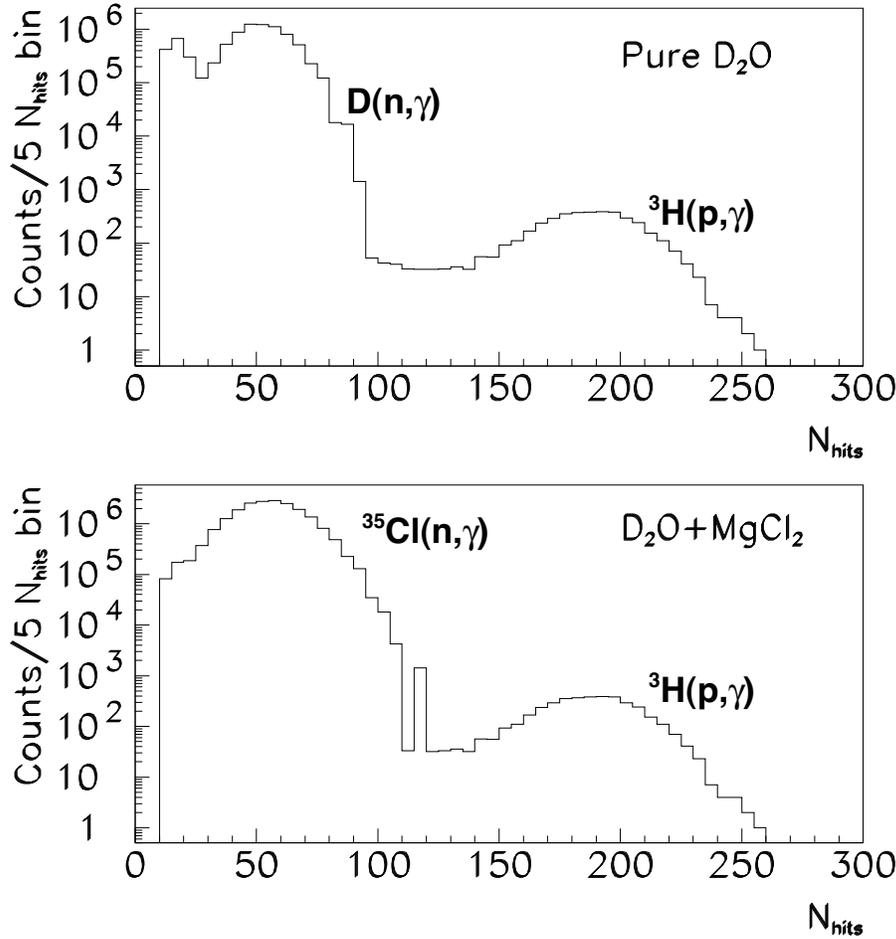} 
	
	\caption{Monte Carlo simulated SNO photomultiplier tube array
	response to neutrons and gamma rays that are generated by the
	\pt\ source.  The abscissa value, $N_{hits}$, is the number of
	photomultiplier tube hits in the SNO detector.  The
	$N_{hits}$-to-energy calibration in this Monte Carlo represents our
	best estimate, but not the calibrated response of the SNO
	detector.  In the pure \hw\ running scenario (top panel), the
	peak centering at $N_{hits}\sim$50 is the 6.25~MeV photopeak
	from $^{2}$H(n,$\gamma$)$^{3}$H. In the salt running scenario,
	neutron capture on $^{35}$Cl generates a gamma cascade with a
	total energy of 8.6~MeV. This is the reason for the broader
	neutron capture peak in the bottom panel.  In these figures, a
	neutron production rate of 2,500~s$^{-1}$ and a gamma-ray
	production rate of 0.6~s$^{-1}$ were assumed.  The sharp
	``peak'' in the bottom panel arises from scaling of the Monte
	Carlo spectrum to correspond to the neutron production rate
	above.  The spectra represent about 3~hours of run time in the
	SNO detector.}
	
	\protect\label{fig:pt_spec_total}  
	\end{center}
\end{figure}


\clearpage

Table~\ref{table:s_fitting}  \\
$\chi^{2}$ minimisation results in fitting the S-factors from 
Hahn~{\it et al.} \\

Table~\ref{table:tritium_gas_composition} \\
Isotopic composition of the tritium gas used in the target. \\

Figure~\ref{fig:source_dwg} \\
Cross sectional drawing of the \pt\ source. \\

Figure~\ref{fig:s_factors} \\
Extrapolated S-factors for the \ptg\ reaction.  Measured S-factors by
Perry and Bame~\cite{perry} and Hahn~{\it et al.}~\cite{hahn} are
shown as data points in this plot.  Hahn~{\it et al.} used a BGO
detector and a NaI detector in their measurements, and the results for
these two detectors are shown separately here.  The solid curve is the
$\chi^{2}$ fitted curve to the combined data in Hahn~{\it et al.} \\

Figure~\ref{fig:estimated_pt_yield} \\
Estimated gamma-ray yield from the \pt\ source.  The yield is plotted
against the mass-1 fraction $f_{1}$ in a 50$\mu$A, 27-keV beam. 
Hydrogen isotopes in the beam and the target were assumed to be
completely mixed.  The yield shown here should be treated as the upper
limit because target degradation was not taken into account in the
calculation.  The dotted lines are the calculated uncertainties based
on the uncertainties in the physical parameters of the constructed
\pt\ source and the cross section (\cite{my_thesis}). \\

Figure~\ref{fig:oht_vacuum_system} \\
Schematic of the scandium tritide target evaporation vacuum system. 
Most of the setup is enclosed in a dry nitrogen environment inside a
glovebox (from~\cite{tritium_workplan}). \\

Figure~\ref{fig:sct2_tritiation} \\
Tritium pumping by the scandium film.  The pumping curve for Doses 1,
5, 9 and 13 are shown here. \\

Figure~\ref{fig:queen_bgo_scheme} \\
Top view of the BGO detector setup for the quality assurance testing of the \pt\ 
source.  The separation between the liquid scintillator (LS) and the target 
of the \pt\ source is about 36~cm. \\

Figure~\ref{fig:baf2_arrangement} \\
Schematic of the BaF$_{2}$ detector system.  The three BaF$_{2}$
detectors were oriented at 45$^{\circ}$ (\detthree), 90$^{\circ}$
(\dettwo), and 135$^{\circ}$ (\detone) to the beam direction, whilst
the neutron detector was oriented at 2$^{\circ}$ to the beam
direction.  The separation between the centre of the target and the
front face of the BaF$_{2}$ crystals was 35.6~cm for \dettwo, and
25.4~cm for \detthree\ and \detone.  The neutron detector was located
at 20.6~cm from the centre of the target. \\

Figure~\ref{fig:n_gamma_sep_tac} \\
Timing distribution of liquid scintillator pulses generated by
neutrons and gamma rays in a $^{9}$Be($\alpha$,n)$^{12}$C$^{\ast}$
source.  Neutrons are cleanly separated from the gamma-rays using the
pulse shape discrimination scheme outlined in the text. \\

Figure~\ref{fig:queens_bgo_mc_fit} \\
Background-subtracted BGO energy spectrum in the quality assurance run
at Queen's University.  The data points constitute the
background-subtracted energy spectrum.  The histogram shown is a fit
using a response function for the BGO spectrometer generated by GEANT.
The measured yield of the \pt\ source during its running at 27~keV is
(0.67$\pm$0.11)~s$^{-1}$.  The excess near 16~MeV was due to
statistical fluctuation, as this was not observed in the later running
of the source.  \\

Figure~\ref{fig:mc_line_shape_compare} \\
Comparing GEANT generated gamma-ray line shape to measurement.  The
data points correspond to the 6.13-MeV line from a calibrated
$^{16}$O$^{\ast}$ de-excitation source.  The solid histogram is the
GEANT generated line shape. \\

Figure~\ref{fig:neutron_generation_scaled} \\
Scaled neutron and gamma-ray production by the \pt\ source at
$E_{p}$=29~keV. The rates were normalised to the current drawn from
the target power supply during the runs.  Also, the production rates
for the $E_{p}$=22~keV and 27~keV runs have been scaled to the
$E_{p}$=29~keV level.  The scaling was done by assuming a pure atomic
beam of protons or tritons since the contribution to the signals from
molecular ions are much smaller.  The ``error bars'' on the
accumulated beam time for the gamma-ray results represent the time
intervals in which the mean production rates were calculated.  The
gamma-ray yield could not be extracted reliably between the ~20-th and
the 50-th hour of the source lifetime because a noise problem in the
electronics system, which was subsequently eliminated.

Figure~\ref{fig:pt_spec_total} \\
Monte Carlo simulated SNO photomultiplier tube array response to
neutrons and gamma rays that are generated by the \pt\ source.  The
abscissa value, $N_{hits}$, is the number of photomultiplier tube hits
in the SNO detector.  The $N_{hits}$-to-energy calibration in this
Monte Carlo represents our best estimate, but not the calibrated
response of the SNO detector.  In the pure \hw\ running scenario (top
panel), the peak centering at $N_{hits}\sim$50 is the 6.25~MeV
photopeak from $^{2}$H(n,$\gamma$)$^{3}$H. In the salt running
scenario, neutron capture on $^{35}$Cl generates a gamma cascade with
a total energy of 8.6~MeV. This is the reason for the broader neutron
capture peak in the bottom panel.  In these figures, a neutron
production rate of 2,500~s$^{-1}$ and a gamma-ray production rate of
0.6~s$^{-1}$ were assumed.  The sharp ``peak'' in the bottom panel
arises from scaling of the Monte Carlo spectrum to correspond to the
neutron production rate above.  The spectra represent about 3~hours of
run time in the SNO detector.

\end{document}